\documentclass[twocolumn,twoside,slac_two]{revtex4}
\usepackage{graphicx}
\usepackage{fancyhdr}
\usepackage{hyperref}
\hypersetup{
    colorlinks=true,
    urlcolor=magenta,
}
\def\Journal#1#2#3#4{{#1} {\bf #2}, #3 (#4)}

\def\PRL{\em Phys. Rev. Lett.}
\def\PRD{{\em Phys. Rev.} D}

\begin{document}

\title{Electron and Positron Fluxes in Primary Cosmic Rays measured with the Alpha Magnetic Spectrometer on the International Space Station}

%

\author{Maura Graziani}
\affiliation{INFN and University of Perugia, I--06100 Perugia, Italy}

\begin{abstract}
Precision measurements by the Alpha Magnetic Spectrometer (AMS-02) on the International Space Station of the primary cosmic-ray electron flux and the positron flux will be discussed. The electron flux and the positron flux each require a description beyond a single power-law spectrum. Both the electron flux and the positron flux change their behavior at $\sim$30 GeV but the fluxes are significantly different in their magnitude and energy dependence. Between 20 and 200 GeV the positron spectral index is significantly harder than the electron spectral index. The determination of the different behavior of the spectral indices versus energy is a new observation and provides important information on the origins of cosmic-ray electrons and positrons.
\end{abstract}

\maketitle

\thispagestyle{fancy}

\section{Introduction}\label{sec:ele_pos_inCR}
The electronic component of CR carries important physics information. Due to their low mass, electrons (e$^{-}$) and positrons (e$^{+}$) are subject to important energy losses in the interaction with the Inter Stellar Medium, during their trajectory between the sources and Earth. For this reason they carry information about the origin and the propagation of CR complementary to the hadronic component. Due to their strong energy losses, electrons and positrons at high energies are unique probes to study the CR source property in the galactic neighborhood \cite{bib:ele}. \\

In 2008, the PAMELA collaboration reported an unexpected \emph{rise} in the positron fraction between 1 and 100\,GeV of energy, which is in contrast with the standard expectations from secondary production models\cite{bib:pamela_1,bib:pamela_2}.
This ``positron excess'' has been confirmed by AMS-02 in the extended energy range of 0.5 - 500\,GeV \cite{bib:Aguilar2014}. Unlike hadrons, light CR leptons are subjected to radiative energy losses which limit the range they can travel to distances of $d\lesssim$\,1\,kpc at $E\gg$\,1\,GeV \cite{bib:ele}. 
Hence the observed positron excess suggests the existence of an unaccounted source of CR leptons, possibly placed near the solar system \cite{bib:Serpico2011}, which manifest itself as a distinctive excess in the high-energy positron flux. Extra sources may include either DM particles annihilation/decay or nearby astrophysical sources such as pulsars or old SNRs \cite{bib:Serpico2011,bib:kopp,bib:DiMauro2014,bib:MertschSarkar2014,bib:TomassettiDonato2015}.\\ 

For a better understanding of the rise in the positron fraction, the measurement of the separate fluxes of electrons and positrons is needed. The measurement of the cosmic electrons (e$^{-}$) and positrons (e$^{+}$) is challenging: charged cosmic rays between 1 GeV - 1 TeV observed at Earth, are made substantially of protons ($\sim90\%$), Helium ($\sim8\%$) and heavy nuclei ($\sim1\%$). e$^{-}$ and e$^{+}$ constitute respectively $\sim$1$\%$ and $\sim 0.1\%$ of the total CR flux. The main challenge in the measurement of the electronic component is the naturally high background/signal ratio. The ratio to the main CR component, i.e. protons (p), amounts to e$^-$/p$\sim10^{-3}$--10$^{-2}$ and e$^{+}$/p $\sim$10$^{-4}$--10$^{-3}$, depending on the energy.
\section{The AMS-02 Detector}
The AMS-02 is a large acceptance cosmic ray detector which has been installed during the STS-134 NASA Endeavour Space Shuttle mission in May 2011 on the International Space Station (ISS), where it will collect cosmic rays until the end of the ISS operation, currently set to 2024. Thanks to the long exposure time combined with a large detector acceptance ($0.5\,$m$^{2}$\,sr), AMS is able to study the primary CR fluxes in the energy range GeV - TeV with unprecedented precision and sensitivity.\\
\begin{figure}[t]
        \centering
        \includegraphics[width=0.35\textwidth]{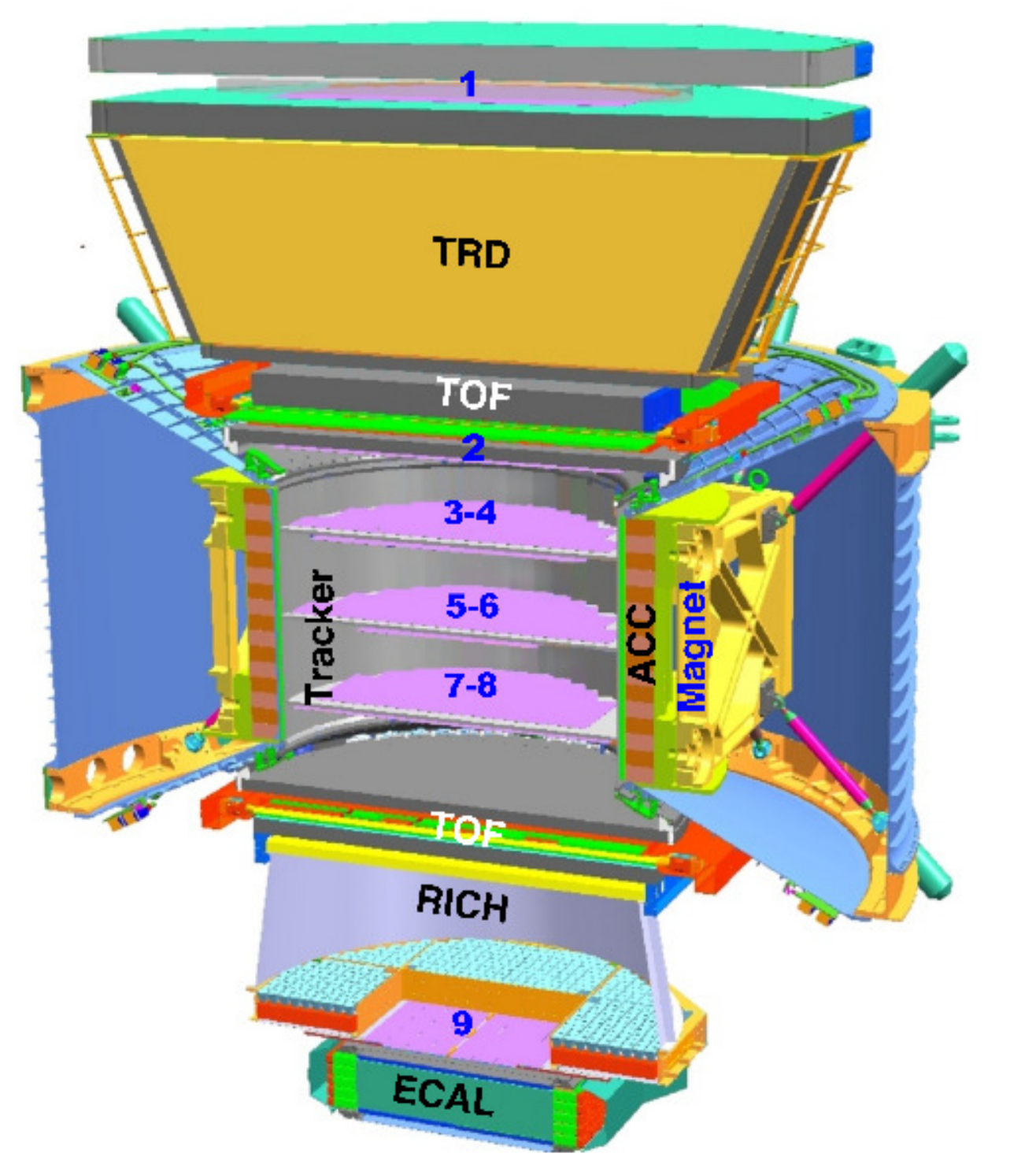}
        \caption{Schematic view of the AMS-02 spectrometer.}
\label{fig:detector}     
\end{figure}

The main goals of the experiment are the direct search of anti-nuclei and indirect search of dark matter particles through their annihilation into standard particles such as $\bar{p}$, $e^{\pm}$, or $\gamma$\--rays \cite{bib:Aguilar2013}. Furthermore, the AMS-02 data are expected to significantly improve our understanding of the CR acceleration and propagation processes in the Galaxy \cite{bib:Oliva2008,bib:Pato2010}. The measurements of low energy CR fluxes during an entire 11 years solar cycle will also help in the understanding of solar physics and of the propagation of CRs in the Heliosphere.\\

The core of the instrument - which schematic layout is reported in Figure \ref{fig:detector} - is a spectrometer, composed of a permanent magnet, which produces a magnetic field with an intensity of 0.14$\,$T, and of a Tracker (TRK) constituted by 9 layers of double-sided micro-strip silicon sensors. The task of the spectrometer is the reconstruction of the trajectory and the measurement of the rigidity ($R=P/eZ$, momentum/charge ratio). Above and below the spectrometer two planes of Time of Flight counters (ToF) provide the main trigger of AMS-02 and distinguish between up-going and down-going particles. This information combined with the trajectory curvature given by the spectrometer, is used to reconstruct the sign of the charge. A Transition Radiation Detector (TRD) is located at the top of the instrument. The detector is completed with a Ring Imaging Cherenkov detector (RICH) and an Electromagnetic CALorimeter (ECAL). The central part of AMS-02 is surrounded by an Anti-Coincidence system (ACC). The AMS-02 detector is described in details in \cite{bib:Aguilar2013}.\\

In the following the AMS's latest results of primary cosmic-ray electron flux and the positron flux will be presented. 
\subsection{The e$^{\pm}$ analysis with AMS-02}
In order to obtaine the needed high e/p rejection power, AMS-02 uses mainly three sub detectors: The TRD, the ECAL and the TRK. TRD and ECAL are the two key detectors for the lepton-hadron separation. These detectors allow the rejection of the overwhelming background coming, mainly, by CR protons, and are used to perform the measurement of the leptonic components.\\

\begin{figure}[t]
        \centering
        \includegraphics[width=0.35\textwidth]{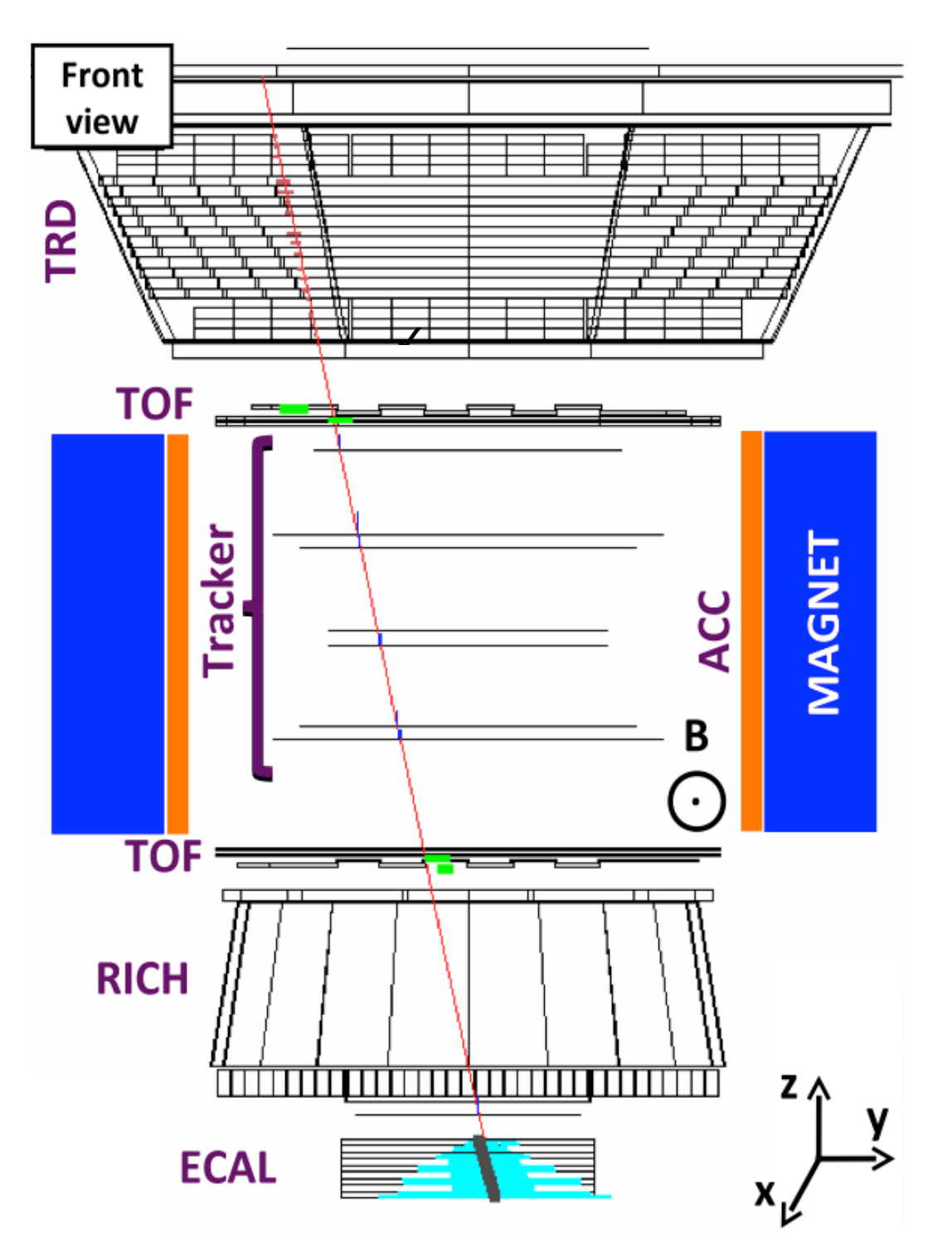}
\caption{A 660\,GeV electron measured by the AMS detector on the ISS in the bending ($y$-$z$) plane. Tracker planes measure the particle charge and momentum. The TRD identifies the particle as an electron. The TOF measures the charge and ensures that the particle is downward-going. The RICH independently measures the charge and velocity. The ECAL,  independently identifies the particle as an electron and measures its energy.}
\label{fig:detector2}     
\end{figure}

The TRK is able to detect the crossing points of particles with high accuracy ($\sim 10\,\mu m$ in the bending direction and $\sim 30\,\mu m$ along the non-bending one, for Z=1 particles). From the crossing points it is possible to provides the measurement of the particle Rigidity. Combining the information about the direction of the particle given by the ToF (down-going or up-going particle) with the sign of the rigidity is possible to reconstruct the sign of the charge and, hence, distinguish between electrons and positrons. An event of 660\,GeV electron measured by the AMS detector combining the information coming from TRK, ECAL and TRD is reported in Fig.\,\ref{fig:detector2}.
\section{Results and conclusions}\label{sec:ele_pos_flux} 
\begin{figure}[t]
\centering
\includegraphics[width=0.5\textwidth]{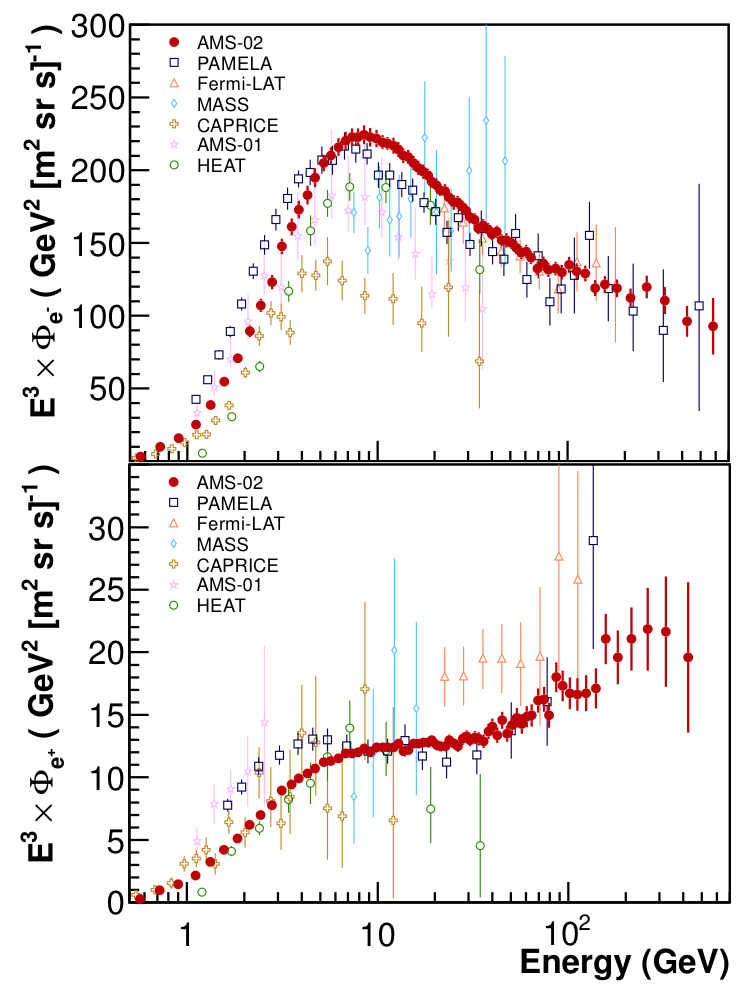}
\caption{Top: The AMS electron flux, multiplied by E$^3$, measured up to 700\,GeV, compared with the earlier measurements. Bottom: the AMS positron flux, multiplied by E$^3$, measured up to 500\,GeV, compared with the earlier measurements.}        
\label{fig:ele_pos_flux}
\end{figure}

\begin{figure}[h!]
\centering
\includegraphics[width=0.5\textwidth]{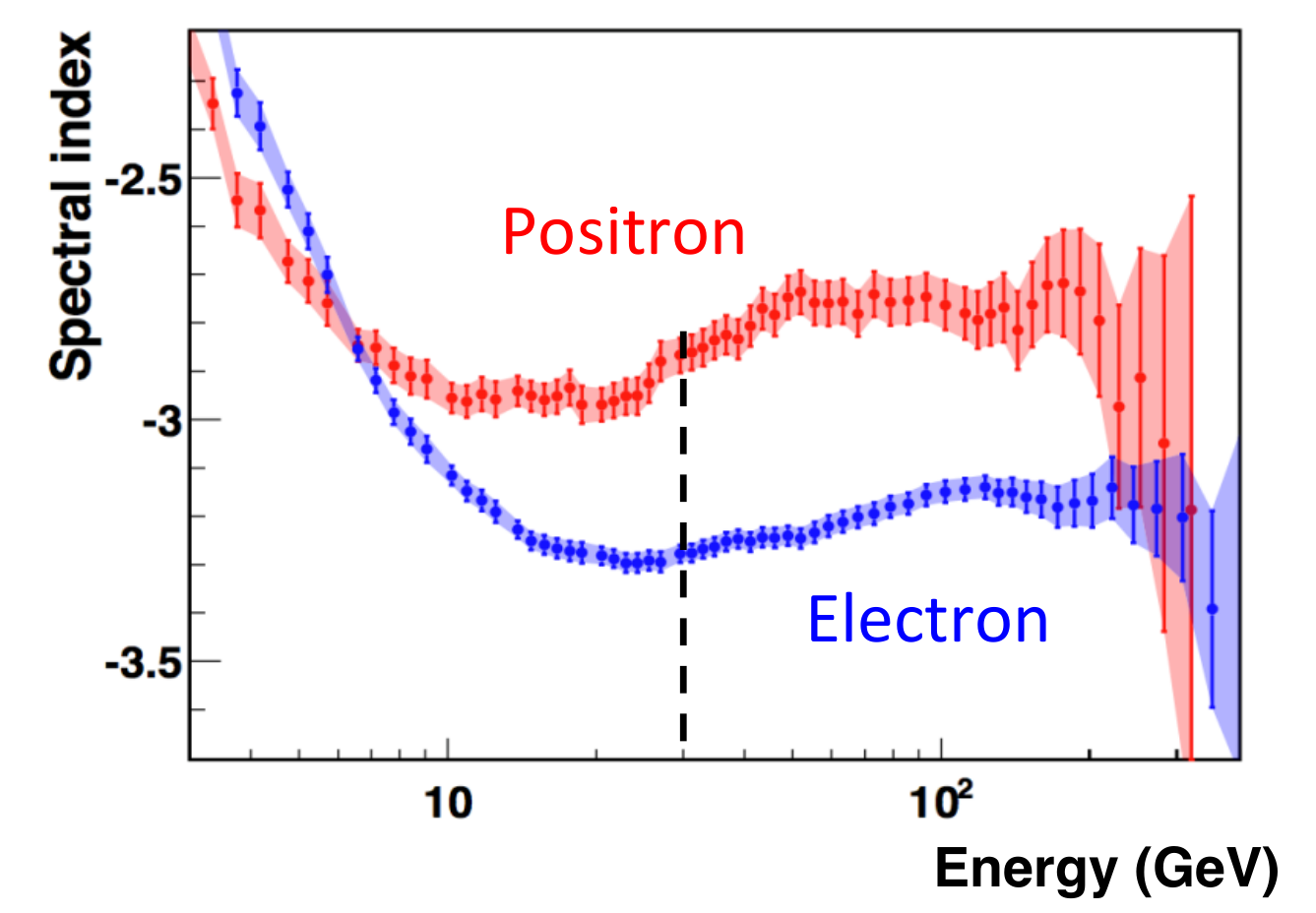}
\caption{The spectral indices of the electron flux (blue points) and
of the positron flux (red points) as a function of energy.}
\label{fig:spectral_index}
\end{figure}

The individual fluxes, multiplied by $\sim E^3$, of $e^{+}$ and $e^{-}$ are shown in the top panels of Fig.\,\ref{fig:ele_pos_flux}. Thank to the high statistic collected by AMS, is possible to determinate the spectral indices as a function of energy as show in Figure \ref{fig:spectral_index}. This is a new observation and provides important information on the origins of cosmic-ray electrons and positrons. The data show that the electron flux and the positron flux each require a description beyond a single power-law spectrum. Both the electron flux and the positron flux change their behavior at $\sim$30 GeV, but the fluxes are significantly different in their magnitude and energy dependence. Above $\sim$\,20 GeV and up to $\sim$\,200 GeV the spectral index of $e^{+}$ flux is significantly harder than the one of $e^{-}$ flux. This demonstrates that the high-energy rise of the positron fraction is due to a hardening of $e^{+}$ spectrum and not to a softening of the $e^{-}$ spectrum. More detailed studies on the $e^{\pm}$ spectral shapes are provided in Ref.\,\cite{bib:ams_ele_pos_fluxes}.\\

The origin of the observed features in the leptonic spectra is currently subjected to extensive phenomenological research. The proposed explanations involve DM annihilation or decay \cite{bib:kopp}, production of $e^{\pm}$ pairs inside nearby pulsars \cite{bib:Serpico2011,bib:DiMauro2014},  or production of $e^{\pm}$ from proton-proton collisions inside old SNRs \cite{bib:MertschSarkar2014,bib:TomassettiDonato2015}. The underlying mechanism can be ascertained by continuing to collect data up to the  TeV energy region and by measuring the $\bar{p}/p$ ratio to high energies \cite{PhysRevLett.117.091103}. \\

\begin{figure}[b]
\centering
\includegraphics[width=0.5\textwidth]{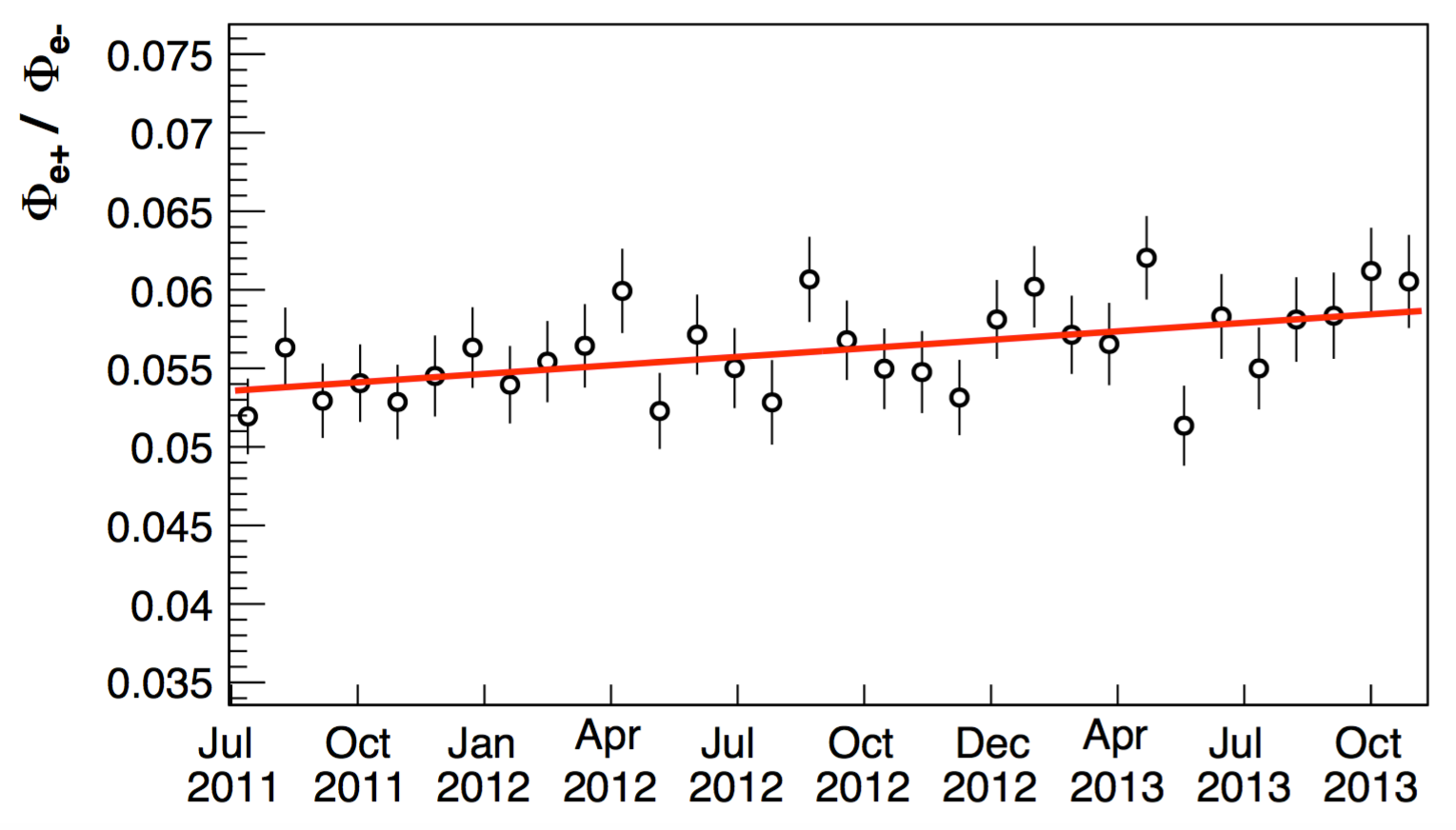}
\caption{Ratio between the positron and electron fluxes as a function of time in one energy bin (4.12-4.54\,GeV). A clear trend with time can be observed.}
\label{fig:ele_pos_in_time}
\end{figure}

It should be also emphasized that the low-energy fluxes are significantly affected by CR transport in the solar wind. In order to interpret the $e^{\pm}$ data at $E\sim$\,0.5-20\,GeV in terms of Galactic CR propagation, the solar modulation effect must be taken into account and properly modeled \cite{bib:pot}. In this respect, AMS-02 has the capability of providing with an essentially continuous monitoring of the $e^{\pm}$ flux evolution with time. Time-variation studies of the CR flux are currently being carried on monthly basis and shorter timescales \cite{bib:tesi_maura}. These studies show different behavior as a function of time for electron and positron fluxes. An example of this feature is reported, in one energy bin (E=4.12 - 4.54\,GeV) as example, in Fig.\ref{fig:ele_pos_in_time} where the ratio between positron and electron flux as a function of time is shown. These data may allow for a substantial progress in the understanding of CR transport in the solar wind. \\

Data recorded in the first $\sim$ 30 months of mission by the AMS experiment have been analyzed and the measurement of the energy spectra of electrons and positrons have been presented. The observed positron excess may imply a heavy Dark Matter WIMP particle or a new mechanism of acceleration in the pulsars. Through the measurement of antiproton to proton ratio and positron fraction to high energies, AMS has the potential to shed a light on the origin of these observed features, either from exotic sources such as dark matter particles or other astrophysical sources such as pulsars.
Furthermore, the simultaneous measurement of e$^{-}$ and e$^{+}$ has also allowed to study differences of the solar modulation effects which are related only to the charge-sign of the particles. These new accurate experimental data will be of great relevance to improve the current models of CR propagation in the heliosphere, therefore allowing a deeper understanding of the local interstellar spectrum of cosmic rays. AMS-02 is foreseen to operate for the entire ISS lifetime, up to 2024, and it will be able to follow in detail the time evolution of the e$^+$ and e$^-$ fluxes along a full solar cycle.
\\
\\
\bigskip 

\end{document}